  \documentclass[acmlarge]{acmart}

\AtBeginDocument{%
  \providecommand\BibTeX{{%
    \normalfont B\kern-0.5em{\scshape i\kern-0.25em b}\kern-0.8em\TeX}}}

\setcopyright{acmcopyright}
\copyrightyear{2025}
\acmYear{2025}
\acmDOI{XXXXXXX.XXXXXXX}

\acmConference[under submission]{under submission}{tbd}{tbd}
\usepackage{rotating}
\usepackage{subcaption}
\usepackage[longtable]{multirow}
\usepackage{tabularx}
\usepackage{enumitem}
\usepackage{xcolor, soul}
\sethlcolor{yellow}

\usepackage{todonotes}
\acmBooktitle{tbd} 
\acmPrice{15.00}
\acmISBN{978-1-4503-XXXX-X/18/06}

\begin{document}

\title[Social MediARverse]{Social MediARverse: Investigating Users' Social Media Content Sharing and Consuming Intentions with Location-Based AR}

\author{Linda Hirsch}
\email{lihirsch@ucsc.edu}
\orcid{0000-0001-7239-7084}
\affiliation{%
  \institution{UC Santa Cruz}
  \city{Santa Cruz}
  \country{USA}
  \postcode{95060}
}
\author{Florian Mueller}
\orcid{0000-0002-9621-6214}
\email{f_m@outlook.com}
\affiliation{%
  \institution{TU Darmstadt}
  \city{Darmstadt}
  \country{Germany}
  \postcode{}
}

\author{Mari Kruse}
\orcid{0000-}
\email{M.Kruse@campus.lmu.de}
\author{Andreas Butz}
\orcid{0000-0002-9007-9888}
\email{butz@ifi.lmu.de}
\affiliation{%
  \institution{LMU Munich}
  \city{Munich}
  \country{Germany}
  \postcode{80539}
}
\author{Robin Welsch}
\orcid{0000-0002-7255-7890}
\email{welschrobin@googlemail.com}
\affiliation{%
  \institution{Aalto University}
  \city{Helsinki}
  \country{Finland}
  \postcode{00076}
}
\renewcommand{\shortauthors}{Hirsch et al.}

\begin{abstract}

Augmented Reality (AR) is evolving to become the next frontier in social media, merging physical and virtual reality into a living metaverse, a Social MediARverse. With this transition, we must understand how different contexts — public, semi-public, and private — affect user engagement with AR content. We address this gap in current research by conducting an online survey with 110 participants, showcasing 36 AR videos, and polling them about the content's fit and appropriateness. Specifically, we manipulated these three spaces, two forms of dynamism (dynamic vs. static), and two dimensionalities (2D vs. 3D). Our findings reveal that dynamic AR content is generally more favorably received than static content. Additionally, users find sharing and engaging with AR content in private settings more comfortable than in others. By this, the study offers valuable insights for designing and implementing future Social MediARverses and guides industry and academia on content visualization and contextual considerations.

\end{abstract}



\begin{teaserfigure}
  \includegraphics[width=\textwidth]{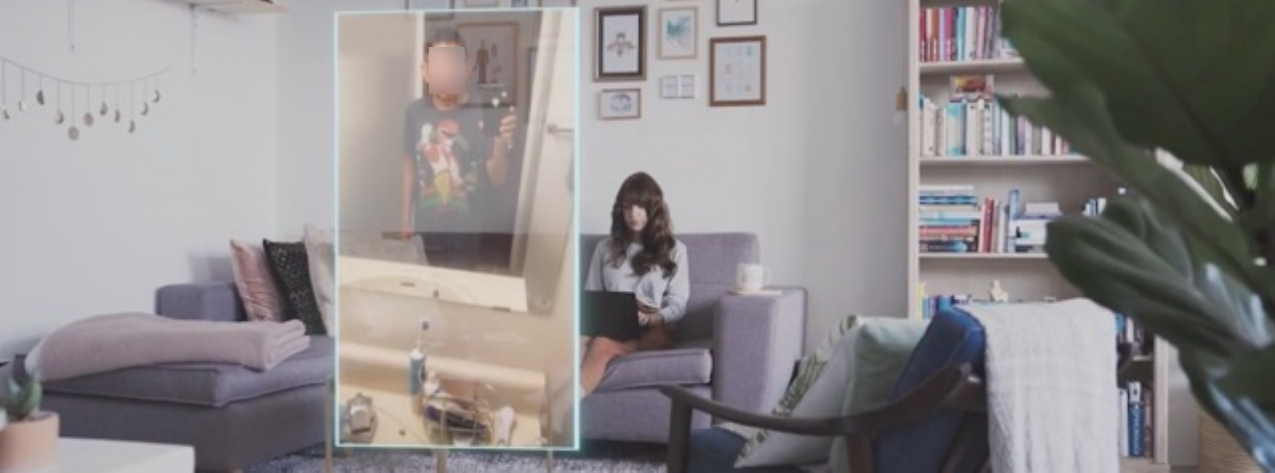}
  \caption{Sharing and consuming Social MediARverse content: The most comfortable space and content visualization format was the private space with embedded dynamic 2D AR content. This example is taken from one of the TikTok videos used in our survey.}
  \Description{This is the teaser figure showing the most preferred content for sharing and consuming in a Social MediARverse. The most comfortable space and content visualization format was combining private space with embedded dynamic 2D AR content. Here, one example is shown along one of the baseline TikTok videos used in the online survey: a dancing man positioned within a living room.}
  \label{fig:teaser}
\end{teaserfigure}

\received{20 February 2007}
\received[revised]{12 March 2009}
\received[accepted]{5 June 2009}

\maketitle

\section{Introduction}
Imagine walking through your streets, wearing your Augmented Reality (AR) glasses, and seeing the AR content others posted into your shared physical environment as they would on social media today. While this might seem like a futuristic imagination, Meta called AR the future technology of social media already in 2017\footnote{\url{https://www.nytimes.com/2017/04/18/technology/mark-zuckerberg-sees-augmented-reality-ecosystem-in-facebook.html}, last accessed August 17, 2023}\footnote{\url{https://www.forbes.com/sites/kathleenchaykowski/2018/03/08/inside-facebooks-bet-on-an-augmented-reality-future/}, last accessed August 17, 2023}\footnote{\url{https://www.vntana.com/blog/augmented-reality-social-media/}, last accessed August 17, 2023} for its characteristic to enhance physical reality with virtual content~\cite{Billinghurst2015}, shifting many interactions with virtual information into the physical world. Through this, information metaphorically breaks through today's screens' limiting and confining glass and enters specific contexts of our physical environments. Such embedded AR content can support community-building~\cite{Silva2017,Kostopoulou2018,Lehto2020} and social interaction~\cite{Hirsch2022_chilbw,Petrov20232,Guo2019} through onsite exploration and content sharing - just as if you walk by your neighbor's house and see what new AR posts they positioned in the front yard. 

Developing AR as the future social media technology merges both into a Social MediARverse toward a ``real-world metaverse''~\cite{Xu2022}. Currently, most shared and consumed social media content is purposefully location-independent to allow immediate and fast communication~\cite{AlQuran_2022}. This works as long as the content is contained in a 2D screen dissociated from its surroundings. Yet, when embedding AR, it becomes part of our 3D physical environment, impacting the places' social and cultural usage. First applications are trying to reinvent location-based AR for social media, such as \textit{Mapstar}\footnote{\url{https://www.mapstar.io/}, last accessed \today} or \textit{Skrite}\footnote{\url{https://www.facebook.com/skriteapp/}, last accessed \today}. Yet, those developments are still far from enabling a real-life Social MediARverse.

Previous work explored shared AR content mainly in and for public spaces~\cite{Nijholt2021} concerning its social acceptability~\cite{Poretski2018,Medeiros2023}, privacy~\cite{Roesner2014,Roesner2014_b,OHagan2023}, or the content creation~\cite{Vera2016,Venta2012}. Whether content and the interaction with it are appropriate and support a good user experience, however, depends on its embedded physical context~\cite{Rico2009,Williamson2011,Dillman2018} and visual display~\cite{Kim2020,Lin2021}. For example, boulders as AR objects in an office space seem unsuitable but fit in an outdoor space~\cite{Dillman2018}. Or, by comparing 2D versus 3D content display in a virtual space, \citet{Kim2020} found that 2D content facilitates information finding, but 3D fosters spatial exploration. Yet, despite these advances, research and industry still lack a systematic understanding of how the physical context and the content design relate and impact users' content-consuming and -sharing intentions. This represents a substantial knowledge gap about location-based AR content-sharing and consumption when transitioning from traditional social media platforms to creating a ubiquitous Social MediARverse.

Our work narrows this knowledge gap by conducting a video-based online survey with N=110 participants, presenting them with 36 AR short-form videos\footnote{All videos are provided here: \url{https://github.com/krufri/SocialMediARverse/tree/main/Videos}.}. In the videos, we vary between three times two times two conditions in a within-subject study design comprising three spatial contexts (private, semi-public, and public spaces) that the content is positioned in, two dimensionalities (2D and 3D), and two dynamics (dynamic versus static content). We identified each condition and its forms through related work, revealing an essential and diverging impact on the user experience deriving from each condition. As they have not been explored in this combination nor the context of a Social MediARverse, our work evaluates their impact on users' comfort level to share and consume it as location-based AR social media content. We tested the conditions by redesigning three TikTok\footnote{\url{https://www.tiktok.com}, last accessed \today} videos embedded into the different spaces. 

Results identify private spaces to bring a significantly higher comfort level for sharing and consuming Social MediARverse content. Additionally, 2D content triggers a significantly higher feeling of comfort, but 3D is significantly less awkward for content consumption across space conditions. Displaying dynamic AR content is more engaging, relatively increases comfort, and lowers arousal than static content. Considering all three conditions, dynamic 2D AR embedded in private space is the most preferred display for Social MediARverse content (see \autoref{fig:teaser}). Consequently, creating a Social MediARverse starts in users' private spaces, embedding the AR content as dynamic 2D short-form videos and acknowledging that usage will change, moving to other forms, such as 3D content. With this, the work contributes to designing future social media networks that use location-based AR toward a ``real-world'' metaverse.

\section{Background and Related Work}
The following section provides a background of current AR social media developments and introduces the purpose and design of location-based AR and current social media trends, including users' sharing and consuming behavior. 

\subsection{Advances in Ubiquitous AR Social Networks} 
Industry and research continuously explore AR for creating ubiquitous social media networks~\cite{Cochrane2016,Braud2022,Rixen2022,Sassmannshausen2021,Hirsch2022_chilbw}. Examples are game apps such as \textit{Pokémon Go}\footnote{\url{https://pokemongolive.com/}, last accessed February 30, 2024} or \textit{Can You See Me Now?}~\cite{Benford2006} that provide social networks fostering a sense of community and motivate spatial explorations. The interaction with the embedded AR content changes the spatial affordance and how users perceive and make meaning of an environment~\cite{Silva2017}. Other work has explored spatially embedded AR to engage younger generations in community matters~\cite{Sassmannshausen2021,Braud2022} or social learning contexts~\cite{Cochrane2016}. For example, \citet{Cochrane2016} developed five AR applications to foster social skills learning by allowing users to geo-tag, negotiate, and share augmented points of interest. Exploring an early version of a location-based AR social network, \citet{Hirsch2022_chilbw}'s show that the interaction with such a network can change the user-place and user-to-user relationships. However, their work also emphasizes the need for privacy settings to moderate who users would want to share their content and locations with. Yet, sharing embedded AR content can also create a feeling of connectedness among strangers~\cite{Petrov20232}. AR social networks can be applied for different purposes and user groups, requiring varying network management settings. The above-mentioned projects explore shared content with a logical or semantic connection between content, user, and environment. This differs from traditional social media content, emphasizing the existence of a research gap when embedding AR social media content into the physical environment. Furthermore, considering that most AR network-related research focuses on location-based AR, we will follow this approach and test the content's appearance in different contexts. 

In addition, prior work identified risks and challenges when transitioning to ubiquitous AR social networks. \citet{Rixen2022} compare users' comfort level when experiencing a person's augmentation through digital content on a smartphone or AR through a head-mounted display while walking the streets. Their findings emphasize that personal AR content triggers lower comfort than digital personal content. Furthermore, \citet{Egtebas2023} discuss the risks of ubiquitous AR social media content being misused for, for example, bullying users or distributing fake news in public spaces. Related to this, \citet{Katell2019} differentiates between private and public spaces considering the varying privacy rights, intimacy levels, and community reach to assess content appropriateness. This emphasizes assessing content appropriateness depending on the spatial contexts and user relationship levels.

\subsection{Purpose and Design Considerations of Location-based AR}
Augmented Reality is used to increase engagement~\cite{Ng2018,Broll2006,Eishita2015,Cisternino2021,Kim2009}, foster social connections and interaction~\cite{Hirsch2022_chilbw,Knoll2023, Cheng2023,Pyae2017,Petrov20232}, or support navigation~\cite{Cheliotis2023,Chatzopoulos2017,Verma2020,Lee2020,sharma2020}. Location-based AR registers content geo-spatially so that users can consume it as embedded AR at the anchored location. Education~\cite{Kleftodimos2023,Li2013}, tourism~\cite{cauchi2019,Spierling2017,Lacka2020}, or community-building~\cite{Kostopoulou2018,Lehto2020,Hirsch2022_chilbw} use location-based AR to foster understanding, meaning-making and relationship building between users and spaces through the spatial and semantic connection of the AR content. Additionally, location-based AR motivates users to explore public outdoor spaces~\cite{Colley2017,Laor2020TheRT} but is also challenged to stay engaging~\cite{Diaz2018}. To achieve this, research suggests making the content more meaningful by relating it to users' lives or fostering social connections~\cite{Hirsch22_embeddedAR,Mekler2019}. 

Prior work has explored the design of AR content in multiple directions. \citet{Yue2019} contribute a location-based AR network enabling users to post and see 3D texts and other 3D models instead of 2D content. Similar to other AR research~\cite{Kim2020,Lin2021}, their work states that 3D is more engaging and fosters more spatial exploration through depth integration and visualization, but flat 2D facilitates information finding~\cite{Kim2020}. The explored content type also differs, including text only, 2D images~\cite{Hirsch2022_chilbw}, 3D objects~\cite{Petrov20232}, audio~\cite{Schroeder2023}, or videos~\cite{Kyza2016}. Design decisions consider the content's engagement character and ability to foster social interaction. Further, \citet{Venta2012} evaluated the information users prefer to consume and share, identifying diverging topics depending on whether it is positioned in private or public spaces. Their work identified that private spaces are more suitable for content related to personal memories, notes, or to-do lists, whereas publicly shared content should be related to local businesses, community activities, or navigation~\cite{Venta2012}. Similarly, \citet{Medeiros2023} compared different places and the social acceptability of interacting with AR content in a public transport situation through a video-based online survey, identifying that location significantly impacts social acceptability. These prior works show that the content's visual dimensionality and spatial context influence the perceived appropriateness and effect of location-based AR content on the user experience and have found great interest in AR research. Our work follows these two prominent conditions as the effect differences have not been explored when using AR for social media communication but can strongly impact the user experience.

\subsection{Sharing and Consuming Social Media Content} 
Social media connect users with their peers~\cite{Nesi2018,statista_purpose_socialmedia}, keep them up-to-date about personal and political topics~\cite{Wang2015, Lee2017}, and can provide a distraction from the everyday~\cite{Wong2017,MATTHES2023107644}. Latest statistics show that image and video content-based social media platforms, such as Instagram\footnote{\url{https://www.instagram.com/}, last accessed \today}, TikTok or YouTube\footnote{\url{https://www.youtube.com}, last accessed \today} are used about 151 minutes per day~\cite{Kemp2023,statista_minutes_socialmedia}. Decisions to create and share content are influenced by how comfortable users feel with the content in the specific context~\cite{SCHOLZ202083,Habib2019}. Videos, particularly short-form videos, are gaining increasing attraction for being easy to access, consume, and create while providing entertainment and short-term breaks from other tasks~\cite{Cheng2023_socialmedia,Liu2021,WANG2020}. A fundamental difference between video and text- and image-based platforms is users' usage intention~\cite{Bartolome2023}: Video-based platforms are mainly used for entertainment, and text- and image-based platforms are used for personal profiling or maintaining social relationships. 

Another influencing factor on social media usage is the content type. Comparing the effect between social media images and short-form videos, \citet{GURTALA2023} find a significant difference in perceived appearance enhancement for images. In another work, \citet{Du2019} compare two 3D social media platforms combining 2D social media content and 3D virtual representations of real-world locations, identifying that 3D representations support more immersive and engaging experiences but fall short regarding the content design realizations. These works provide first insights into the effect of 2D and 3D and static images versus dynamic video social media content but do not reveal any insights about the conditions' level of appropriateness in comparison or depending on the context. Our work considers the previous works' findings regarding different content dynamism and dimensionality effects and similarly explores its effect on user comfort and usage intentions when presented as Social MediARverse content in this study.



\subsection{Knowledge Gap and Research Questions}
Content appropriateness and design regarding different spaces, dimensionalities, or dynamics has shown to strongly influence the user experience of either location-based AR or social media content. Therefore, we consider these dimensions relevant when aiming to create a Social MediARverse. However, research lacks the knowledge of how they impact each other and users' sharing and consuming intentions.
In our work, we narrow this gap while being guided by the following research questions:
\begin{itemize}
    \item [\textbf{RQ1}] How do the surrounding space, dimensionality, and dynamics affect users' comfort in sharing and consuming AR social media content?
    \item [\textbf{RQ2}] What meaning does the relationship between space, dimensionality, and dynamics relationship have for transitioning from digital social media to a ubiquitous Social MediARverse?
\end{itemize}

We consider these questions relevant because the results will allow us to provide directions for future researchers and social network designers and what they need to consider for creating the next Social MediARverse that users will be comfortable using. We approach these questions using short-form videos because those are currently the most engaging type of shared social media content\footnote{\url{https://www.forbes.com/advisor/business/social-media-statistics/\#source}, last accessed August 27, 2024}. To be more precise, we consider three TikTok videos because the platform has over a billion users and uses short-form videos (about 3-60 seconds) as a communication medium~\cite{GURTALA2023} and has been used in previous HCI (Human-Computer Interaction) studies~\cite{Chiossi2023}.

\begin{figure*}
    \includegraphics[width=\textwidth]{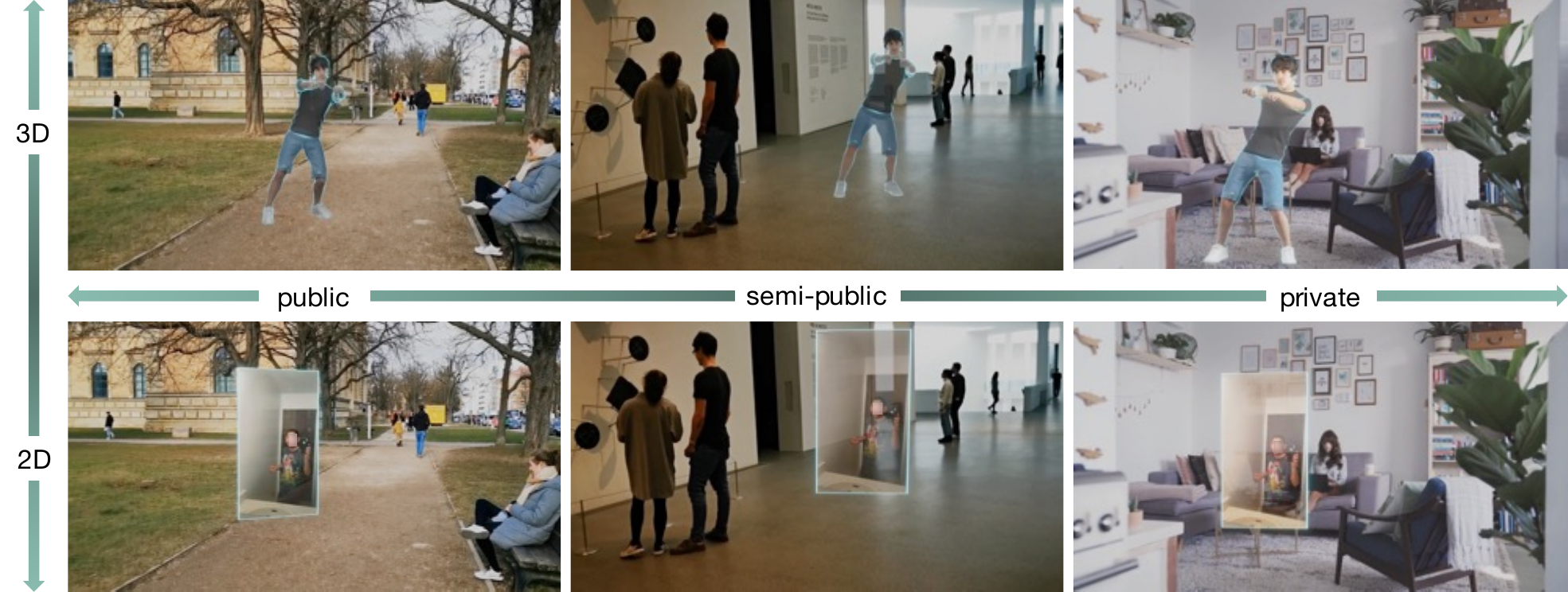}
  \caption{Space and Dimensionality in our videos: Public, semi-public, and private combined with the 2D and 3D dimensionality. The dynamic condition cannot be visualized through static images but is included in the supplementary video. }
  \Description{The figure shows the implementation of the different space and dimensionality dimension combinations in six images.}
  \label{fig:space_dimensionality_images}
\end{figure*}

\section{Methodology}
We approached the gap with a video-based online survey using Qualtrics\footnote{\url{https://www.qualtrics.com}, last accessed \today} for survey creation and data collection. While online surveys cannot represent a real-world encounter or external influences, they allow for a larger and more diverse reach~\cite{Koelle2020}. Furthermore, video-based web surveys are established and viable methods to compare the effect of different contexts that facilitate participants to imagine the observed interaction as realistic scenarios~\cite{Koelle2020,Medeiros2023}. To gain broad insights into this new field of research and to identify generalizable design indicators instead of qualitative in-depth understanding~\cite{creswell2003research}, we decided on this approach with a focus on the quantitative data.

\subsection{Independent Variables}
\label{sec:meth:ivs}
We defined three independent variables: dimensionality (2D vs. 3D), space (private, semi-public, public), and dynamics (static vs. dynamic), resulting in $2\times3\times2 = 12$ conditions. \autoref{fig:space_dimensionality_images} shows the realization of the space and dimensionality conditions. For increasing validity, we tested the conditions in the three most liked TikTok videos according to Wikipedia\footnote{\url{https://web.archive.org/web/20230412032600/https://en.wikipedia.org/wiki/List_of_most-liked_TikTok_videos}, last accessed February 2nd, 2023.}, showing a dancing man\footnote{\url{https://www.tiktok.com/@jamie32bsh/video/7058186727248235782}, last accessed September 12, 2023, {\fontencoding{TS1}\selectfont\symbol{"A9}} Jamie Big Sorrel Horse.}, lip-syncing\footnote{\url{https://www.tiktok.com/@bellapoarch/video/6862153058223197445}, last accessed September 12, 2023, {\fontencoding{TS1}\selectfont\symbol{"A9}} Bella Poarch.}, and a drawing video\footnote{\url{https://www.tiktok.com/@fredziownik_art/video/6911406868699073798}, last accessed September 12, 2023, {\fontencoding{TS1}\selectfont\symbol{"A9}} Franek Bielak.}, which resulted in a total of $12\times3 = 36$ trials per participant.
 In this work, we define a private space as a space accessible only for and managed by certain people~\cite{Birch2010} that can be individualized and personalized~\cite{Cooke1999}, semi-public as privately owned spaces accessible by a certain structural group formed by, e.g., common interests or shared activities~\cite{Orhan2022}, and public space as a physically and socially accessible space by everyone that is centrally managed by the city or communal authority~\cite{Rahman2014,Miller2007}.

\subsection{Dependent Variables Reflected in the Survey} 
As our dependent variables, we researched the appropriateness level by assessing the different content displays and contexts' effects on participants' emotional responses using the SAM questionnaire~\cite{BRADLEY199449} and their comfort level for sharing and consuming content in the given scenario. Based on prior work~\cite{Lukoff2018}, we considered the sharing behavior as \textit{active} and consuming as \textit{passive} social media usage. We asked about participants' comfort level using a 7-point Likert scale with five self-developed statements, similar to \citet{Habib2019}: 1) \textit{I feel comfortable seeing the displayed content}, 2) \textit{I feel awkward seeing this content in this space}, 3) \textit{I would like to see more of this content}, 4) \textit{I would feel comfortable placing the displayed content myself}. 5) \textit{Placing this content would leave me feel akward about myself and what others think about me}. 
Further, we ask about the preferred audience by considering different types of relationships based on social media settings similar to 
Facebook\footnote{\url{https://www.facebook.com/help/211513702214269?helpref=faq_content}, last accessed \today.} and as applied by previous work~\cite{Hirsch2022_chilbw}.

\subsection{Additional Survey Questions}
In addition, we collected qualitative statements through open-ended and control questions about participants' current social media usage and demographics. Please consider the supplementary material for further details on the survey.

\paragraph{Open-ended Survey Questions}
We added three questions with open-ended comment fields to learn about participants' reasoning for their choices. This included a question about their reasons for selecting a particular content type most suitable for location-based AR and their motivation to consume and create location-based AR social media content. We also asked participants to indicate where they would feel comfortable sharing this AR content for each video. 

\paragraph{Control Questions} 
We asked participants about their regular social media usage and prior AR experience. If they were not using social media, they were automatically excluded. Further, participants rated the similarity between the original TikTok video and the 3D visualizations on a 0-100\% scale to identify potential issues in the video creation after having watched all the videos.

\paragraph{Demographics}
We collected gender, age, highest educational degree, nationality, and primary occupation.

\subsection{Video Creation}
We created the videos in Blender\footnote{\url{https://www.blender.org/features/video-editing/}, last accessed \today}, blurring the TikTok logo and creator aliases for the study to reduce negative or positive sentiment towards the platform. To implement the dynamics, we selected the videos (dynamic, 2D) and their screenshots (static, 2D).
For the 3D content, we created static and animated 3D objects. We found one model from the library Mixamo\footnote{\url{https://www.mixamo.com/}, last accessed August 20, 2023} and self-created the other two. For the space dimension, we shot two background videos and downloaded one from the free video library Pexels \footnote{\url{https://www.pexels.com/}, by Taryn Elliot, last accessed August 20, 2023}. All videos included a slow forward motion as if the watcher, or, in our case, the study participants watching the videos from a first-person perspective, was passing by the AR content on foot. By this, we used the motion parallax - the perception that an object's changed position due to an altered viewing perspective~\cite{SHERMAN200375} - to increase participants' depth perception and, thus, their sense of 3D and reality similarly to previous AR research~\cite{Knoerlein2007,Furmanski2002}. Additionally, the same people (who consented) were visible in the public and semi-public conditions to increase the feeling of being in a shared environment. Further, we increased transparency, added an emission shader to the visual borders of the content to enhance the holographic and digital AR characteristics, and added the original TikTok soundtracks. The resulting 36 videos are each ten seconds long and in 16:9 format. The format and participants' hands-free video consumption support our anticipation of content interaction via AR glasses.

\subsection{Procedure}
We pilot-tested the survey with two external researchers before publishing. Afterward, we distributed the survey via Prolific\footnote{\url{https://www.prolific.com/}, last accessed \today} screening for English-speaking and social media-using participants. We informed them about the project goal and data privacy settings following GDPR. With their consent, they accessed the main survey, which presented a short explanation about AR being used in ``prospective social networks'' and a definition of the different space types\footnote{Instruction given in the survey: ``Now, we present you 36 videos where augmented reality content is placed in real-world spaces. These spaces can be either private (like a living room or kitchen), semi-public (like museums or movie theaters), or public (like streets or squares). As you watch the videos, \textbf{imagine yourself in these locations} and how you would \textbf{feel seeing this AR content} as social media content shared by other users.'' Similarly for sharing: ``Imagine you yourself would post and \textbf{place this AR content} in a private space. With whom would you \textbf{share }the displayed content in the given scenario, and therefore would be able to see it?''}. Then, participants watched the videos in a randomized order. For each video, we asked participants to imagine being users of the Social MediARverse and, first, passively consuming the AR content at the presented location. In a second step, we asked them to imagine actively sharing and positioning the content on social media in this space. By this, we gradually increased participants' responsibility for the content - from passive consumption to active sharing. After completion, we compensated participants with nine euros. The project was approved by the first author's university's ethics board.

\subsection{Control Study: Terminology} We followed the main study up with a control study (n=30) about the impact of our phrasing of the task description. We compared the three task description conditions: 1) \textit{Main study} as in the original study, 2) mentioning \textit{wearing AR glasses}, or 3) that the seen content was \textit{publicly shared}. We did not find meaningful differences in the ratings. The order was randomized and also conducted using qualtrics and prolific. We provide results in the attachment in \autoref{tab:textConditionControl}.

\subsection{Participants}
We invited 113 people to our survey.  Three participants completed the survey exceptionally fast (below 25 minutes; the rest completed the survey with 55 minutes), which were excluded. The remaining 110 participants produced complete data sets. All participants consume social media content at least once a week (n = 77 more than two hours daily, n = 30 daily but less than two hours, n = 2 every two days, and n = 1 once a week). 
On average, participants were $M=29$ years old ($SD = 8.84$). 58 participants identified as female, 50 identified as male, one person reported being non-binary/third gender, and one person preferred not to disclose their gender identity. The nationalities of the subjects varied geographically. The most frequently represented nationalities were South African (n = 31), Polish (n = 26), and Portuguese (n = 10), in addition to twelve other nationalities (n = 43). 89 participants had prior experience with AR applications such as Pokémon GO, Google Lens, Live View, or Snapchat filters, 21 had no prior experience and were not skeptical about viewing AR content but showed privacy concerns when sharing it. Most (n = 100) consume social media videos on platforms like YouTube and Instagram. 

\section{Results}

\begin{figure*}[t!]
	
	\centering
	\includegraphics[width=\textwidth]{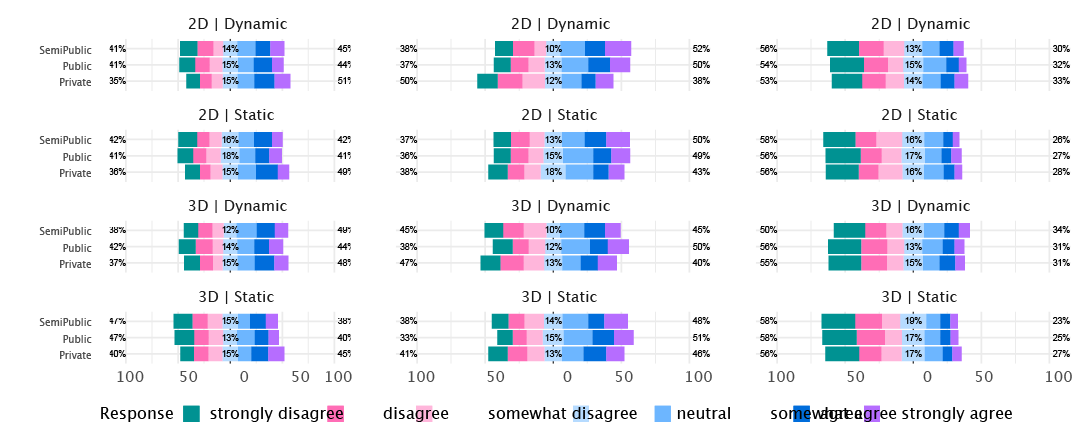}\hfill
	\vspace{-1em}
	\begin{minipage}[t]{.33\linewidth}
		\centering
		\subcaption{Comfortable Seeing}\label{fig:results:comfortable_seeing}
	\end{minipage}%
	\begin{minipage}[t]{.33\linewidth}
		\centering
		\subcaption{Awkward Seeing}\label{fig:results:awkward_seeing}
	\end{minipage}%
	\begin{minipage}[t]{.33\linewidth}
		\centering
		\subcaption{Like to See More}\label{fig:results:like_to_see_more}
	\end{minipage}%

	\caption{Likert scale results for the statements 
"\textit{I feel \textbf{comfortable seeing }the displayed content}",
"\textit{I feel \textbf{awkward seeing }this content in this space}", and
and "\textit{I would \textbf{like to see more} of this content}".}
	\Description[short]{The figure shows the Likert scale results and raw data distribution for the items \textit{\textbf{Comfortable Seeing}}, \textit{\textbf{Awkward Seeing}}, and \textit{\textbf{Like to See More}}.}	
	\label{fig:likert_1}
\end{figure*}
\begin{figure*}[t!]
	
	\centering
	\includegraphics[width=\textwidth]{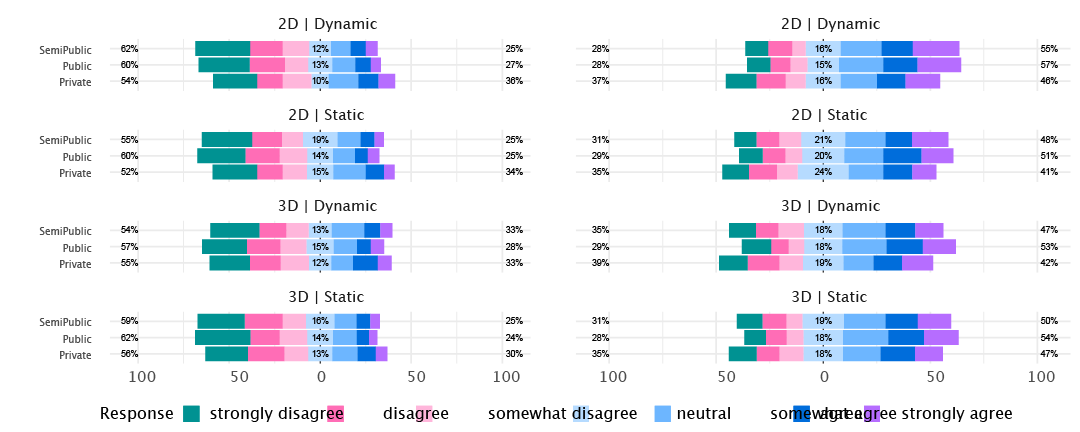}\hfill
	\vspace{-1em}
	\begin{minipage}[t]{.5\linewidth}
		\centering
		\subcaption{Comfortable Placing}\label{fig:results:comfortable_placing}
	\end{minipage}%
	\begin{minipage}[t]{.5\linewidth}
		\centering
		\subcaption{Awkward Placing}\label{fig:results:awkward_placing}
	\end{minipage}%
	
	\caption{Likert scale results for the statements "\textit{I would feel \textbf{comfortable placing} the displayed content myself}" and "\textit{\textbf{Placing} this content would leave me feel \textbf{awkward} about myself and what others think about me}".}
	\Description[short]{The figure shows the Likert scale results and raw data distribution for the items \textit{\textbf{Comfortable Placing}}, and \textit{\textbf{Awkward Placing}}.}	
	\label{fig:likert_2}
\end{figure*}
\begin{figure*}[t!]
	
	\centering
	\includegraphics[width=\textwidth]{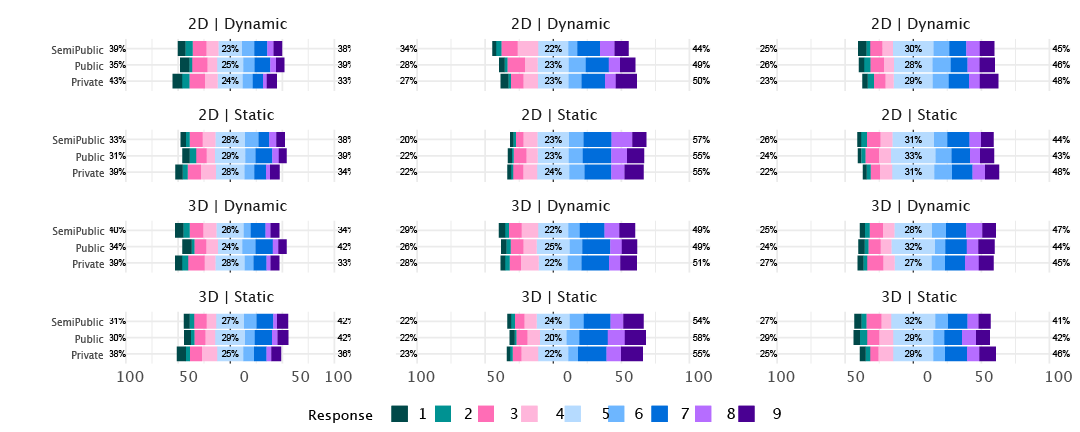}\hfill
	\vspace{-1em}
	\begin{minipage}[t]{.33\linewidth}
		\centering
		\subcaption{Valence}\label{fig:results:sam_valence}
	\end{minipage}%
	\begin{minipage}[t]{.33\linewidth}
		\centering
		\subcaption{Arousal}\label{fig:results:sam_arousal}
	\end{minipage}%
	\begin{minipage}[t]{.33\linewidth}
		\centering
		\subcaption{Dominance}\label{fig:results:sam_dominance}
	\end{minipage}%

	\caption{The SAM results split into Valence, Arousal, and Dominance for each condition combination.}
	\Description[short]{The figure shows the SAM results split into Valence, Arousal, and Dominance and grouped by conditions. Similar to the Likert scale plots, the SAM also shows the raw data distribution for each of the 2x2x3 conditions.}	
	\label{fig:sam}
\end{figure*}
\begin{figure*}[t!]
	
	\centering
	\includegraphics[width=\textwidth]{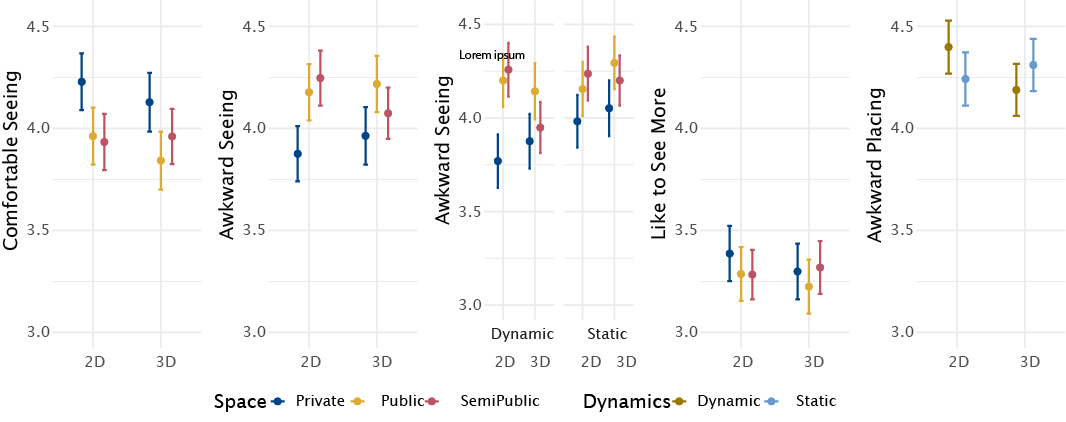}\hfill
	\vspace{-1em}
	\begin{minipage}[t]{.19\linewidth}
		\centering
		\subcaption{Comfortable Seeing\\Dimensionality:Space}\label{fig:results:interaction:comfortable_seeing_dim_space}
	\end{minipage}
	\begin{minipage}[t]{.19\linewidth}
	\centering
	\subcaption{Awkward Seeing\\Dimensionality:Space\\}\label{fig:results:awkward_seeing_dim_space}
	\end{minipage}
	\begin{minipage}[t]{.19\linewidth}
	\centering
	\subcaption{Awkward Seeing\\Dynamics:Dim.:Space}\label{fig:results:awkward_seeing_dim_space_dyn}
	\end{minipage}
	\begin{minipage}[t]{.19\linewidth}
	\centering
	\subcaption{Like to See More\\Dimensionality:Space}\label{fig:results:like_to_see_dimensionality_space}
	\end{minipage}
	\begin{minipage}[t]{.19\linewidth}
	\centering
	\subcaption{Awkward Placing\\Dynamics:Dimensionality}\label{fig:results:interaction:awkward_placing_dyn_dim}
	\end{minipage}

	\caption{The significant interaction effects between independent variables.}
	\Description[short]{The figure highlights the significant interaction effects between independent variables.}	
	\label{fig:interactions}
\end{figure*}

We applied descriptive and inferential statistics to evaluate the quantitative data. Complementary, we applied two types of analysis for the qualitative feedback: content analysis and thematic analysis. To analyze our ordinal data, we used the ART package and computed a linear mixed model (LMM) for each rank-aligned dependent variable with space, dynamics, and dimensionality as a predictor and participant as a random effect term with random slopes for each TikTok video. Therefore, each participant's individual effect of the video displayed is controlled. All post-hoc tests concerning more than two variables were corrected with the Bonferroni method.
For the results of all models, see \autoref{tab:sam} and \autoref{tab:questionnaire}. Regarding the similarity between original TikTok and 3D representations, the dancing man (see \autoref{fig:teaser}) received lowest similarity ratings ($M$=29\%, $SSTD$=26\%), followed by the lip-syncing with $M$=33\% ($STD$=28\%), and the drawing video with $M$= 53\% (28\%). For the sake of brevity, we will only discuss significant effects in more detail. However, all our conditions' mean values can be found in \autoref{tab:Meanssam} and \autoref{tab:Meanscomfort}. 

\begin{table}[ht]
\centering
\caption{Means (and standard deviations) for dynamics, dimensionality, space, valence, arousal, and dominance.}
  \label{tab:Meanssam}
  \Description{The table shows the overview of the descriptive SAM results across all dimensions.}
\small
\begin{tabularx}{0.7\textwidth}{lccllc}
  \toprule
\textbf{ Dynamics} & \textbf{Dimensionality} & \textbf{Space} & \textbf{Valence} & \textbf{Arousal} &\textbf{ Dominance} \\ 
  \midrule
 Dynamic & 2D & Private & 4.8 (2.27) & 5.72 (2.24) & 5.67 (2.15) \\
 Dynamic & 2D & Public & 5.07 (2.19) & 5.59 (2.1) & 5.53 (2.11) \\
 Dynamic & 2D & Semi-public & 5 (2.22) & 5.52 (2.12) & 5.51 (2.15) \\
 Dynamic & 3D & Private & 4.92 (2.14) & 5.71 (2.1) & 5.56 (2.13) \\
 Dynamic & 3D & Public & 5.16 (2.18) & 5.67 (2.1) & 5.59 (2.11) \\
 Dynamic & 3D & Semi-public & 4.94 (2.2) & 5.66 (2.16) & 5.62 (2.09) \\
 Static & 2D & Private & 4.98 (2.14) & 5.95 (2.02) & 5.69 (1.95) \\
 Static & 2D & Public & 5.15 (2.15) & 5.94 (2.04) & 5.55 (1.96) \\
 Static & 2D & Semi-public & 5.18 (2.05) & 6.02 (1.95) & 5.52 (2) \\
 Static & 3D & Private & 5.02 (2.2) & 6.05 (2.07) & 5.67 (2.09) \\
 Static & 3D & Public & 5.34 (2.15) & 6.07 (2.12) & 5.43 (2.15) \\
 Static & 3D & Semi-public & 5.29 (2.13) & 5.95 (2.07) & 5.37 (2.08) \\
  \bottomrule
\end{tabularx}
\end{table}

\begin{table}[ht]
\centering
\caption{Means (and standard deviations) for comfort measures across dynamics, dimensionality, and space.}
  \label{tab:Meanscomfort}
    \Description{The Table presents the overview of all comfort means and standard deviations (in brackets).}
\small
\begin{tabularx}{\textwidth}{lccXXXXX}
  \toprule
\textbf{ Dynamics} & \textbf{Dimensionality} & \textbf{Space} & \textbf{Comfortable Seeing} & \textbf{Awkward Seeing} &\textbf{Like to See More} & \textbf{Comfortable Placing} & \textbf{Awkward Placing }\\ 
  \midrule
 Dynamic & 2D & Private & 4.28 (1.99) & 3.77 (1.98) & 3.49 (2) & 3.49 (2.02) & 4.12 (2.02) \\
 Dynamic & 2D & Public & 3.98 (1.97) & 4.2 (1.97) & 3.31 (1.9) & 3.11 (1.89) & 4.54 (1.96) \\
 Dynamic & 2D & Semi-public & 3.96 (2.04) & 4.26 (2.08) & 3.35 (1.93) & 3.06 (1.93) & 4.53 (1.98) \\
 Dynamic & 3D & Private & 4.14 (2.02) & 3.88 (2.01) & 3.35 (1.95) & 3.46 (1.97) & 4.04 (1.97) \\
 Dynamic & 3D & Public & 3.94 (2.04) & 4.14 (2) & 3.31 (1.94) & 3.25 (1.9) & 4.36 (1.94) \\
 Dynamic & 3D & Semi-public & 4.13 (1.98) & 3.95 (1.96) & 3.49 (1.97) & 3.33 (1.96) & 4.17 (1.89) \\
 Static & 2D & Private & 4.18 (1.94) & 3.98 (1.89) & 3.28 (1.83) & 3.4 (1.91) & 4.04 (1.86) \\
 Static & 2D & Public & 3.95 (1.94) & 4.15 (1.93) & 3.27 (1.91) & 3.13 (1.87) & 4.37 (1.9) \\
 Static & 2D & Semi-public & 3.9 (1.99) & 4.24 (2.03) & 3.22 (1.78) & 3.16 (1.86) & 4.32 (1.89) \\
 Static & 3D & Private & 4.12 (2.01) & 4.05 (2) & 3.25 (1.87) & 3.3 (1.91) & 4.17 (1.92) \\
 Static & 3D & Public & 3.75 (1.98) & 4.29 (1.89) & 3.14 (1.85) & 3.02 (1.84) & 4.45 (1.87) \\
 Static & 3D & Semi-public & 3.79 (2.02) & 4.2 (1.97) & 3.15 (1.82) & 3.11 (1.84) & 4.31 (1.92) \\
  \bottomrule
\end{tabularx}
\end{table}

\subsection{SAM-Assessment: Valence, Arousal, Dominance} 
All  SAM ratings averaged across stimuli and participants can be found in \autoref{tab:Meanssam} and \autoref{fig:sam}. Analysis with the LMM highlighted two main effects on valence; see \autoref{tab:sam}. There was higher valence, meaning more negative, in the static content condition ($M$ = 5.16, $SD$ = 1.35) as compared to the dynamic condition ($M = 4.98$, $SD = 1.37$). For space, we calculated post-hoc contrasts. The valence rating for the private space ($M = 4.93$, $SD = 1.45$) was significantly lower than that for the public space ($M = 5.18$, $SD = 1.44$), $z = -5.764$, $p < .001$. Similarly, the valence rating for the private space was also significantly lower than the semi-public space ($M = 5.10$, $SD = 1.34$) $z = -3.475$, $p = .002$. The difference between the public and semi-public spaces was not statistically significant, $z = 2.289$, $p = .066$.

Only dynamics revealed a significant main effect for arousal, \autoref{tab:sam}. Dynamic content produces less arousal ($M = 5.64$, $SD = 1.41$) compared to static visualizations ($M = 6.00$, $SD = 1.28$); therefore, dynamic visualizations are calmer than static ones. Our manipulations had no effects on dominance ratings or the dimensionality overall; see \autoref{tab:Meanssam}.


\subsection{Comfort Ratings} 
We asked about the \textit{comfort} of sharing or consuming the visualization across all experimental conditions (see \autoref{fig:likert_1} and \autoref{fig:likert_2}). On the item \textit{I feel \textbf{comfortable seeing }the displayed content}, we found an effect of dynamics, see \autoref{tab:questionnaire},  with higher ratings for dynamic ($M = 4.07$, $SD = 1.33$) as compared to static content ($M = 3.95$, $SD = 1.37$). We also found a main effect for dimensionality with 2D ($M = 4.04$, $SD = 1.35$) content being more comfortable than 3D content ($M = 3.98$, $SD = 1.37$). Comfort ratings also differed based on the type of space (\autoref{tab:questionnaire}). The content was rated as more comfortable in the private space ($M = 4.18$, $SD = 1.42$) compared to both the public space ($M = 3.90$, $SD = 1.42$) and the semi-public space ($M = 3.95$, $SD = 1.38$). Specifically, post-hoc contrasts revealed a significant difference between private and public spaces, with a difference of 194.5, $SE = 28.5$, $z = 6.832$, $p < .001$. Similarly, there was a significant difference between private and semi-public spaces, with a difference of 132.0, $SE = 28.5$, $z = 4.638$, $p < .001$. However, the difference in comfort ratings between the public and semi-public spaces was not statistically significant, $z=-2.194$, $p = .084$. 

We investigated the space $\times$ dimensionality interaction by comparing the effect of space across both dimensionalities. In the 2D dimensionality, there were significant differences among the conditions. Specifically, the contrast between the private and public conditions was significant, $z = -5.057$, $p < .001$. However, the difference between the private and semi-public conditions was not significant, $z = -1.777$, $p = .227$. Importantly, the public and semi-public conditions differed significantly, $z = 3.280$, $p = .003$. For the 3D dimensionality, the contrast between the private and public conditions was significant, $z = -3.030$, $p = .007$. Additionally, a notable difference was observed between the private and semi-public conditions, $z = -4.095$, $p < .001$. However, the difference between the public and semi-public conditions was not statistically significant, $z = -1.065$, $p = .861$. Overall, these results indicate that the effects of space were more pronounced in the 2D compared to the 3D dimensionality. Further details can be found in \autoref{fig:interactions}.

For the item \textit{I feel \textbf{awkward seeing} this content in this space}, we find a very similar pattern of results. There was a main effect of dynamics (dynamic: $M = 4.03$, $SD = 1.25$ vs. static: $M = 4.15$, $SD = 1.29$). There was also a main effect of space with private ($M = 3.92$, $SD = 1.37$) being significantly lower than public ($M = 4.20$, $SD = 1.39$; $z = -6.19$, $p < .001$) and semi-public ($M = 4.16$, $SD = 1.31$; $z = -3.93$, $p < .001$). Again, this was qualified by a space $\times$ dimensionality interaction mirroring the previous results. For the 2D dimensionality, the contrast between the private and public conditions was significant, $z = 3.190$, $p = .004$, while the difference between the private and semi-public conditions was not, $z = -0.157$, $p = 1.0$. The public and semi-public conditions differed significantly, $z = -3.348$, $p = .002$. In the 3D dimensionality, both the contrasts between the private and public, $z = 3.315$, $p = .003$, and the private and semi-public conditions were significant, $z = 4.701$, $ p <.001$. However, the public and semi-public contrast was not, $z = 1.386$ (\autoref{fig:interactions}). We analyzed the three-way interaction by comparing the effect of space for each combination of dimensionality and dynamics. We found only an effect of space for static conditions, all other $p>.05$, when comparing private and semi-public in 2D,  $z = 3.810$, $p <.001$, and for 3D when comparing private and public,  $z = 3.315$, $p = .003$; 

\begin{table}[htbp]
	\centering
	\caption{The RM ANOVA results for all dimensions and combinations for the SAM questions.}
	\label{tab:sam}
	\Description{The Table shows the RM ANOVA results for all dimensions and combinations for the SAM questions.}
	\begin{tabular}{lccccl}
		\toprule
		\textbf{Dependent Variable} & \textbf{Model Term} & \textbf{F} & \textbf{\textit{p}} & \textbf{Sig} \\
		\midrule
		\multirow{7}{*}{\textbf{Valence}} & Dynamics & 12.74 & < .001 & *** \\
		& Dimensionality & 1.21 & .272 & \\
		& Space & 16.84 & < .001 & *** \\
		& Dynamics:Dimensionality & 0.61 & .435 & \\
		& Dynamics:Space & 0.14 & .866 & \\
		& Dimensionality:Space & 0.40 & .669 & \\
		& Dynamics:Dimensionality:Space & 0.75 & .473 & \\
		\midrule
		\multirow{7}{*}{\textbf{Arousal}} & Dynamics & 61.70 & < .001 & *** \\
		& Dimensionality & 0.79 & .375 & \\
		& Space & 2.07 & .126 & \\
		& Dynamics:Dimensionality & 2.52 & .113 & \\
		& Dynamics:Space & 0.95 & .387 & \\
		& Dimensionality:Space & 2.06 & .127 & \\
		& Dynamics:Dimensionality:Space & 1.41 & .245 & \\
		\midrule
		\multirow{7}{*}{\textbf{Dominance}} & Dynamics & 1.51 & .219 & \\
		& Dimensionality & 0.73 & .391 & \\
		& Space & 1.09 & .335 & \\
		& Dynamics:Dimensionality & 0.80 & .371 & \\
		& Dynamics:Space & 0.27 & .766 & \\
		& Dimensionality:Space & 1.43 & .240 & \\
		& Dynamics:Dimensionality:Space & 0.26 & .773 & \\
		\bottomrule
	\end{tabular}
\end{table}
\begin{table}[htbp]
	\centering
	\caption{The RM ANOVA results for all dimensions and combinations for the comfort Likert scale questions.}
	\label{tab:questionnaire}
	\Description{The Table shows the RM ANOVA results for all dimensions and combinations for the comfort Likert scale questions.}
	\begin{tabular}{lcccl}
		\toprule
		\textbf{Dependent Variable} & \textbf{Model Term} & \textbf{F} & \textbf{\textit{p}} & \textbf{sig} \\
		\midrule
		\multirow{7}{*}{\textbf{Comfortable Seeing}} & Dynamics & 6.28 & .012 & * \\
		& Dimensionality & 4.75 & .029 & * \\
		& Space & 24.34 & < .001 & *** \\
		& Dynamics:Dimensionality & 2.04 & .153 & \\
		& Dynamics:Space & 0.88 & .416 & \\
		& Dimensionality:Space & 4.73 & .009 & ** \\
		& Dynamics:Dimensionality:Space & .58 & .577 & \\
		\midrule
		\multirow{7}{*}{\textbf{Awkward Seeing}} & Dynamics & 16.78 & < .001 & *** \\
		& Dimensionality & 0.06 & .800 & \\
		& Space & 19.65 & < .001 & *** \\
		& Dynamics:Dimensionality & 3.00 & .083 & \\
		& Dynamics:Space & 2.99 & .050 & \\
		& Dimensionality:Space & 7.67 & < .001 & *** \\
		& Dynamics:Dimensionality:Space & 3.14 & .044 & * \\
		\midrule
		\multirow{7}{*}{\textbf{Like to See More}} & Dynamics & 15.88 & < .001 & *** \\
		& Dimensionality & 1.01 & .315 & \\
		& Space & 6.51 & .002 & ** \\
		& Dynamics:Dimensionality & 0.77 & .379 & \\
		& Dynamics:Space & 2.54 & .079 & \\
		& Dimensionality:Space & 4.17 & .016 & * \\
		& Dynamics:Dimensionality:Space & 1.00 & .367 & \\
		\midrule
		\multirow{7}{*}{\textbf{Comfortable Placing}} & Dynamics & 6.59 & .010 & * \\
		& Dimensionality & 0.27 & .612 & \\
		& Space & 14.44 & < .001 & *** \\
		& Dynamics:Dimensionality & 2.50 & .114 & \\
		& Dynamics:Space & 0.36 & .694 & \\
		& Dimensionality:Space & 1.10 & .333 & \\
		& Dynamics:Dimensionality:Space & 0.89 & .410 & \\
		\midrule
		\multirow{7}{*}{\textbf{Awkward Placing}} & Dynamics & 0.51 & .475 & \\
		& Dimensionality & 1.34 & .247 & \\
		& Space & 41.96 & < .001 & *** \\
		& Dynamics:Dimensionality & 17.81 & < .001 & *** \\
		& Dynamics:Space & 0.59 & .552 & \\
		& Dimensionality:Space & 1.39 & .250 & \\
		& Dynamics:Dimensionality:Space & 0.03 & .968 & \\
		\bottomrule
	\end{tabular}
\end{table}

On the item \textit{I would \textbf{like to see more} of this content}, 
participants exhibited a main effect of dynamics and space. Again, participants rated that they would like to see more of this content for the dynamic ($M = 3.38$, $SD = 1.30$) compared to the static condition ($M = 3.22$, $SD = 1.29$). Corrected post-hoc tests indicated that private ($M = 3.34$, $SD = 1.37$) differed from public ($M = 3.26$, $SD = 1.34$;  $z = -3.39$, $p < .001$) and public from semi-public ($M = 3.30$, $SD = 1.27$;  $z = -2.747$, $p = .018$). This differed again as a function of dimensionality (space $\times$ dimensionality), see also \autoref{fig:interactions}. Post-hoc analyses on 2D revealed significant differences for private and public conditions, $z = -2.529$, $p = .034$. However, the difference between private and semi-public conditions was not statistically significant, $z = 0.842$, $p = 1.0$. A significant difference emerged between the public and semi-public conditions, $z = 3.371$, $p = .002$. For 3D, the contrast between private and public conditions was significant, $z = -2.481$, $p = .023$. There was also a significant difference between private and semi-public conditions, $z = -2.670$, $p = .023$. However, the contrast between public and semi-public conditions did not yield a significant difference, $z = -0.189$, $p=1.0$.

Due to convergence issues, the analysis on \textit{I would feel \textbf{comfortable placing} the displayed content myself}
had to be conducted without random slopes for each stimulus. We found a main effect of dynamics (dynamic: $M = 3.28$, $SD = 1.27$ vs. static: $M = 3.19$, $SD = 1.28$) and of space. Participants were more comfortable sharing content themselves in private spaces ($M = 3.41$, $SD = 1.38$) as compared to public ($M = 3.13$, $SD = 1.36$;  $z = 4.90$, $p < .001$) and semi-public space ($M = 3.17$, $SD = 1.27$;  $z = 4.36$, $p < .001$). None of the higher-order interactions were significant, all $p >.05$. 

For \textit{\textbf{Placing} this content would leave me feel \textbf{awkward} about myself and what others think about me}, we find only a main effect of space with private space being less awkward ($M = 4.09$, $SD = 1.38$)  than semi-public ($M = 4.33$, $SD = 1.31$;  $z = -5.83$, $p < .001$) and public space ($M = 4.43$, $SD = 1.40$;  $z = -9.03$, $p < .001$) but also semi-public and public differed significantly from each other ($z = 3.20$, $p =.004$). Again, the space $\times$ dimensionality interaction could be qualified by testing 2D and 3D separately, see \autoref{fig:interactions}. For 2D, the private vs. public contrast was significant, $z = 3.74$, $p < .001$. However, the private vs. semi-public ($z=1.697$, $p=.269$) and public vs. semi-public ($z=-2.039$, $p=.125$) contrasts were not significant, respectively.
For 3D, both the private vs. public and private vs. semi-public contrasts were significant, $z =3.41$, $p = .002$, and $z = 3.55$, $p < .001$, respectively. The public vs. semi-public contrast was not significant, $z = 0.15$, $p=1.0$.

\subsection{Content Analysis: Alternative Space Suggestions}
The content analysis was applied to code and count the mentioned spaces where participants would feel comfortable sharing the video content. According to \citet{Vaismoradi2013}, content analysis is suitable for quantifying textual statements, which was our main interest for this question's responses. We introduced three codes, \textbf{public space}, \textbf{semi-public space}, and \textbf{private space} in the codebook (see \autoref{tab:codebook} in the attachment) to identify patterns, following the definitions as presented in section \ref{sec:meth:ivs}. Two researchers coded the data manually and independently before comparing the results following the approach by \citet{Neuendorf2017}. 

Participants made alternative suggestions in all but one case (the private static 3D video). 
The content analysis resulted in an agreement rate of 92\% and interrater reliability of $\kappa$ = 0.86 using Cohen's Kappa, indicating 'almost perfect' reliability~\cite{Landis1977}. Main differences related to handling unspecific answers, such as ``Indoors''. The results showed that participants would feel comfortable sharing the video content mainly in private contexts (139 counts) and suggested semi (32)- or public (17) spaces only a few times in comparison. 

\subsection{Thematic Analysis: Suitability and Usage of Location-Based AR Social Media}
Lastly, we report the results about what content participants find most suitable for a Social MediARverse, why, and what they would like to consume and share via such a medium. For this, two authors used an inductive approach, conducting a reflexive thematic analysis for the qualitative feedback. Following \citet{Braun2019}'s approach, we 1) familiarized ourselves with the data, 2) noted down initial codes, 3) clustered the codes in the first themes, 4) iterated on those themes, 5) finalized theme definitions in an agreement between both researchers and 6) summarized the findings. Additionally, we report participants' preferences regarding content visualization and space with absolute numbers. Qualitative results are kept short for a limited amount of data and keep the focus on the quantitative results.

\subsubsection{Suitability of Content} Participants found dynamic 2D AR as the most suitable (n=42) content type, followed by animated 3D objects (n=34), 3D objects (n=22), images (n=9), and none (n=3). In their explanations, we identified three themes, \textit{Realistic}, \textit{Entertainment}, \textit{Integration in the Environment} with 15 codes in total. The \textit{Realistic} theme represents two perspectives, one concerning 3D content (dynamic and static) to be more adapted to its physical surroundings, e.g., P80 stating \textit{``Because it would fit to 3D environment as I am not looking at screen.''} The second perspective relates to dynamic 2D content, the most known and familiar format considering common short-form social media videos. Additionally, the 2D content origin seems traceable to a real, other user, allowing participants to draw relations to their own life; e.g., P54, \textit{``For me it's because it's most similar to real life and everyday activities that we see.''} In comparison, 3D content was often linked to games and not a real person, complicating feeling connected to or identifying with it.

The \textit{Entertainment} theme solely concerned dynamic content, 2D and 3D. Participants found the dynamic 3D content ``fun'', ``futuristic'', and ``creative'', emphasizing the novelty of 3D AR social media content. Furthermore, they appreciated the movement in the dynamic 2D and 3D videos and emphasized feeling more engaged and entertained by dynamic content, e.g., P10 about dynamic 2D content: \textit{``It is more entertaining and grabs attention more than the other options.''} 3D content was further perceived as less personal and more creative due to its novelty and less realistic visual appearance.
In the \textit{Integration in the Environment} theme, the content's appropriateness for the respective space and its positioning were relevant factors to define the content's suitability. This included the consideration of other people in the shared space or passers-by and what they would like to see, e.g., P23 about dynamic 3D \textit{``It will engage more people to interact.''} The theme relates to all content types and revealed concerns about potential clutter and distraction resulting from particularly dynamic content and in the nine times preferred 2D static content in the ranking.

\subsubsection{Content to Share and Consume} 44 participants indicated an interest in using such a technology, whereas 52 would rather not use it, and 14 stayed indecisive. The thematic analysis resulted in five themes related to the motivation of (not) using a Social MediARverse, \textit{Entertainment}, \textit{Profession or Education}, \textit{Staying up to date}, \textit{Location-Specific}, \textit{Non-Use}, including 14 codes. The \textit{Entertainment} theme revealed that participants appreciated the creativity and interactivity of a Social MediARverse. They mentioned benefits regarding extending the social network, newly experiencing a familiar environment, and a novel way for self-expression (e.g., P67 \textit{``I'd use it just as I use normal social media, it'd be another form of expression of my feelings, myself, etc.''}). Related to participants' \textit{Profession or Education}, they would like to consume tutorials or distribute advertisements. Under the \textit{Staying up to date} theme, we clustered general news or content posted by friends, family, or members of the same community. Furthermore, the theme \textit{Location-Specific} comprises information about travel, history, navigation, or local events. Participants also differentiated between spaces. In private spaces, they preferred seeing content by friends and family and smaller objects such as flowers or pets. In comparison, in semi-public and public spaces, the content should relate to the space's purpose, e.g., seeing the drawing process of an exhibition piece in an art gallery. Yet, participants also raised safety and privacy concerns, leading to a tendency not to use location-based AR social media (\textit{Non-Use}). The main concern derived from sharing personal location data with strangers. This theme emphasizes securing users' locations from shared, location-based content and against potential misuse.

\section{Discussion}
Our work explores the influence of different \textit{spaces} for content embedding and content designs (\textit{dimensionality} and \textit{dynamics}) of location-based AR on users' tendency to share and consume social media content (\textbf{RQ1}). The most suitable content design and space combination is the dynamic 2D AR content embedded in private space (see \autoref{fig:teaser}). Further, we identified the meaning of those dimensions and relationships for transitioning from digital social media to a ubiquitous Social MediARverse (\textbf{RQ2}), which we will discuss below.

\subsection{Appropriate AR Content Design Considering Space, Dimensionality, and Dynamics}
Our results revealed multiple significant effects between the conditions, resulting in the following takeaways:

\paragraph{The familiarity of 2D content design trumps 3D for consumption but not for sharing.}  Our results reveal that 2D dynamic content is the preferred content design to consume because it resembles traditional social media content and realism. This is at odds with prior work, finding that 3D was preferred for triggering greater engagement and realistic character representation~\cite{Yue2019,Kim2020}. However, \citet{Kim2020}'s results derive from a virtual 3D environment interaction, whereas \citet{Yue2019} did not compare generated content to known social media content designs. We assume that our content's short-form video social media character sets users' expectations of how to get content presented. Yet, we also expect it to change when augmented 3D content becomes more normal and ubiquitous. For now, 2D content is more recognizable as social media content, further supporting trust toward the content's origin than 3D.

\paragraph{Private spaces are the most comfortable environments for sharing and consuming Social MediARverse content.} Our results identify private spaces as the most comfortable and suitable environment for sharing and consuming content. Thereby, the content displayed in private may be reduced to posts by friends, family, and other community members, which introduces a more personal and intimate relation to the content and its owners. It also aligns with the purposes of using social media for maintaining social relationships as identified by, e.g., ~\citet{statista_purpose_socialmedia}. Based on \citet{Birch2010}, we hypothesize that the emphasis on private spaces derives from the feeling of ownership, perceived control, and greater intimacy. Implementing a Social MediARverse in private spaces would enable users to revisit videos from past private events or post content that might interest another person later. At the same time, our findings emphasize that users are more concerned about sharing their content in physical spaces than in the digital social media realm, considering that users often share their profiles online with ``everybody''~\cite{Burkell2014}. Yet, by extending location-based AR into private spaces, we expect it can support users in feeling more ``at home'' if used to foster social exchange and reminisce. The finding might change when introducing privacy settings in our contexts but will require further exploration and negotiating of physical and digital space ownership~\cite{Katell2019}. Either way, it increases the complexity when designing a Social MediARverse with location-based AR, as we need to develop settings that account for the different combinations of the digital and physical. This includes private spaces where the perceived feeling of safety is greater than in other spaces but the digital connections can invade and harm it.


\paragraph{Sharing and consuming intentions significantly differ between spaces and dimensionality, emphasizing greater sensitivity for public places.}
Users found consuming and placing 3D content in private spaces significantly less awkward than in the other two spaces but were more open toward 2D content. There, semi-public places were perceived as similar to private spaces. The finding indicates a greater sensitivity to sharing and consuming content in public than in the other spaces for both dimensionalities and even stronger for 3D content. We assume the difference derives from the 3D design's gamified and futuristic appearance and entertainment character, which might increase its perceived inappropriateness for semi-public and public spaces but still be perceived as suitable for private spaces. In semi-public and public, we are also more exposed to others' opinions and potential public shaming~\cite{Warren2010}. To mitigate uncomfortable user experiences, we suggest limiting specifically 3D to space-fitting content and only supporting a free choice of dimensionality and content in private spaces.

\paragraph{Dynamic content is more engaging and comfortable to share and consume privately than static AR content.} 
Our findings indicate dynamic content is preferred over static content in private for being more engaging, attention-grabbing, and entertaining. This aligns with prior work regarding the consumption of images vs. video content on traditional social media~\cite{GURTALA2023} and extends it by providing insights about users' sharing intentions and preferences in dependence on the space. Additionally, public spaces can be quite attention-demanding~\cite{Foth2018,Kiss2019}. We assume that quieter private spaces are more suitable for dynamic content because they have fewer distractions. However, the dynamics in our study videos were also spatially limited and excluded any interfering external movements, e.g., crossing cyclists or cars. Balancing the AR content's movement radius depending on the physical space, the users' position, and the video content might support sharing dynamic content in semi-public and public and optimize its consumption in private.


\subsection{Creating a Social MediARverse: Design Considerations, Concerns and Outstanding Questions}
Our work further provides insights into new design potentials and concerns related to a ubiquitous Social MediARverse. 


\subsubsection{Content Management According to the Space and Space Ownership} 
Current social media practices already consider users' location to improve content recommendation~\cite{Joe2021,Farseev2017,Kukka2017} and manage content ownership. Our work emphasizes the same need for a Social MediARverse depending on the space type. To manage shared content, we suggest classifying the spaces (e.g., through a meaning of place framework~\cite{GUSTAFSON20015}) to provide either 2D or 3D content respectively, and reviewing it regarding the space's socio-cultural norms to assure appropriateness~\cite{Egtebas2023} and suitability for semi-public and public spaces. This is further impacted by who the content is shared with and the consideration of passers-by's perspective~\cite{OHagan2023}, which we did not address in this work. In comparison, managing content anchored in private spaces should involve people living in that space and be relatable to users on a personal and social level. This raises further research questions about the content's lifetime - e.g., should it depend on the person's stay in the space or the person's lifetime? Who can share content, and do people have to be on-site to post? Or how can content be reviewed for appropriateness depending on the context but independent of who shared it? 

\subsubsection{Location-Sharing Issues in Light of Private Space Preference} In addition to customizing the content design according to the space type, our qualitative results revealed privacy concerns when sharing users' locations. This results in a paradox because private households were the preferred yet are the more personal and intimate~\cite{Cooke1999} spaces. This finding aligns with the social media privacy paradox related to users raising concerns about handing over ownership of personal data to third parties but doing so freely and willingly at the same time~\cite{norberg2007,Zafeiropoulou2013}. Location sharing increases the risk of experiencing physical or property harm~\cite{Kostakos2011,Tsai2009}, which shows our participants' concerns relevancy. Potential countermeasures relate to notifying users when others request their location data~\cite{Kostakos2011} or anonymizing location-based content for users outside the creator's network~\cite{Jones2011}, emphasizing the need for content management in a Social MediARverse. However, our work does not reveal further location-based AR-specific risks other than well-known privacy challenges with location-based content. Thus, future work must explore potential AR-specific risks that have not been explored yet.

\subsubsection{Dynamic Short-Form Videos in 2D and 3D}
Another design consideration concerns providing dynamic content for users to share and consume. In line with prior work~\cite{GURTALA2023}, our results support that dynamic content is also more suitable for a Social MediARverse if the intention is to provide users with entertainment~\cite{statista_purpose_socialmedia} similar to original TikTok videos~\cite{Bartolome2023}. While dynamic 2D is perceived as more appropriate overall, dynamic 3D supports a more futuristic and creative content display that embeds better into the 3D physical surroundings. By this, 2D and 3D trigger different associations supporting diverging interaction experiences. Dynamic 3D should be considered for game networks or entertainment platforms with less intimate personal connections, whereas dynamic 2D seems promising to support personal relationships and users in connecting the content to their own lives. 
Part of this might be caused by current social media content being predominantly created by personal 2D recording devices (i.e., mobile phone cameras). In comparison, 3D technologies are only available to more professional content creators. At the same time, we also see potential in combining 2D and 3D in the same post in future developments. Yet, for now, a Social MediARverse should focus on dynamic content, and its designers should decide between 2D and 3D considering its purpose.

\subsubsection{The Level of Realism to Represent Humans in AR } Current work embedding social media content in urban spaces via AR or in Virtual Reality (VR) mainly restrains to 2D icons and posts in a 3D environment~\cite{Kukka2017}. While this choice of 2D displays aligns with our results, we expect 3D AR visualizations to become more realistic~\cite{Haller2004} and, thus, more relatable and suitable in the future. At the same time, we agree with \citet{Slater2020} about the risks of too realistic 3D AR, considering users might struggle to recognize what is AR and what is ``real''. The challenge increases with AR glasses when users forget they are wearing them. Further, \citet{Huang2022} showed that virtual human representations cause users to keep a certain physical distance, avoiding entering personal space. Consequently, 3D AR models of realistically appearing humans will impact users' navigating physical spaces. In contrast, social media content must satisfy a certain level of realism and recognizability, mainly related to authentically displayed humans~\cite{Joshi2023,Jun2020}, to build trust between content creators and consumers, e.g., influencers and followers. This is a similar challenge emphasized in prior VR work and the lifelike resemblance of virtual avatars, where a greater likeliness could lead to higher engagement~\cite{Moustafe2018}. Thus, for creating a Social MediARverse, research should explore balancing users' lifelike 3D representations to enable (self-)identification under considering VR avatar design findings while avoiding too gamified or unrelated designs.

\section{Limitations and Future Work}
Our work faces limitations regarding the study design and scope. For one, we conducted an online survey, which allows testing for multiple conditions with farther reach. 
Yet, it limits the potential of reflecting real-world scenarios and cannot fully reflect a real-world experience. The results might change by actually wearing an AR headset and walking through the different spaces. Thus, future work should evaluate our findings in real-life spaces and within an actual Social MediARverse network, also considering risks such as cluttering~\cite{Colley2017} and other negative impacts of short-form videos~\cite{Chiossi2023}. Second, we mainly gathered quantitative results to identify generalizable opportunities and challenges. Complementary to our work and the abovementioned point, we suggest future work to approach the topic qualitatively to collect an in-depth understanding of the indicated choices and real-world influences.
Third, we based our content on short-form videos, a special social media content mainly used for entertainment. This ignores other content types and their representation as AR content, such as pure texts or emoticons. In future projects, the content types should be extended and assessed in comparison for the comfort level and emotional response, similar to work by \citet{Kukka2017} testing social media icons in a 3D virtual city but for AR.
Lastly, we used the three most liked TikTok videos to create the AR content. Each TikTok content already affects the user experience. While we counter-balanced the effect by considering three videos, the content is thematically very similar. The effect might differ for other thematic content, such as more serious, political content. However, the range of topics and their representation in short-form videos is huge, so we selected videos from a publicly available list. Yet, this shows a research opportunity to explore potential differences according to the content theme. 

\section{Conclusion}
The work contributes to the future of creating ubiquitous social networks using location-based AR by exploring the effect of different content dimensionality (2D and 3D), dynamics (dynamic and static), and spaces (public, semi-public, and private) on users' sharing and consuming intentions. Our results highlight private spaces as environments to embed, share, and consume Social MediARverse content. Furthermore, 3D content is more engaging, but 2D is preferred altogether. The most prominent content design is dynamic 2D content embedded in private space. Our work contributes insights into users' sharing and consuming preferences and emphasizes future research potential for creating a ubiquitous, ``real-world'' metaverse. 
\section{Acknowledgments}
\begin{acks}
Stays empty for anonymization purposes...
\end{acks}

\bibliographystyle{ACM-Reference-Format}
\bibliography{socialmediaverse_reference}

\section{Appendices}
\subsection{The Codebook}

\begin{table}[ht]
    \centering
    \caption{The Codebook: We coded the additional suggested spaces by participants, differentiating between three codes: public, semi-public, and private space. The codes focus on urban spaces.}
        \label{tab:codebook}
        \Description{The Codebook: We coded the additional suggested spaces by participants, differentiating between three codes: public, semi-public, and private space. The codes focus on urban spaces.}
    \begin{tabular}{l|c}
    \textbf{Code} & \textbf{Description}\\
    \toprule
        \multirow{1}{*}{Public space} &  \parbox{12cm} {Physically and socially accessible by everyone, centrally managed by the city or communal authority~\cite{Rahman2014,Miller2007}. Public spaces incorporate a highly relevant socio-cultural role for society and the local communities by providing the space to meet, exchange, and develop\footnote{\url{https://unhabitat.org/topic/public-space}, last accessed \today}.}\\
        &\\
        \multirow{1}{*}{Semi-public space} &  \parbox{12cm} {Privately owned and accessible by a certain structural group formed by, e.g., common interests or shared activities~\cite{Orhan2022}.}\\
        &\\
        \multirow{1}{*}{Private space} & \parbox{12cm} {Accessible only for and managed by certain people permitted and defined by law~\cite{Birch2010}. A space to individualize and personalize, including self-defined rules~\cite{Cooke1999}.}\\
    \bottomrule
    \end{tabular}
\end{table}

\subsection{Control Study: Task Description}

\begin{table}[ht]
    \centering
    \caption{The table shows the mean and standard deviation (SD) for the three conditions describing the task: \textit{Wearing AR Glasses}, \textit{Public}, \textit{Main Study}.}
    \begin{tabular}{l|c|c|c}
        Question & Wearing AR Glasses & Public & Main Study Wording \\
         \toprule
        Comfortable Seeing. & 3.53 (1.84)  & 3.33 (1.91) & 3.4 (1.89) \\
        Awkward Seeing. & 4.59 (1.92) & 4.92 (1.92) & 4.72 (1.92)\\
        Like to See More. & 2.99 (1.83) & 2.74 (1.77)& 2.91 (1.78)\\
        Comfortable Placing. & 3.13 (1.83) & 2.88 (1.85) & 3.08 (1.79) \\
        Awkward Placing. &4.58 (1.96)& 4.79 (1.99)& 4.45 (2) \\
        Valence (SAM) & 5.19 (2.11)& 5.33 (2.04)& 5.31 (2.12)\\
        Arousal (SAM) & 5.34 (2.12) & 5.34 (2.18) & 5.41 (2.27)\\
        Dominance (SAM) &5.06 (2.12) & 4.97 (2.18)& 5.01 (2.27) \\
         \bottomrule
    \end{tabular}
    \label{tab:textConditionControl}
\end{table}

\end{document}